\documentstyle[aps,preprint]{revtex}

\begin{document}

\widetext

\draft

\title{Flow induced ultrasound scattering: experimental studies.}

\author{Sh. Seifer and V. Steinberg}
\address{Department of Physics of Complex Systems,
  Weizmann Institute of Science, Rehovot, 76100, Israel}
\date{\today}

\maketitle
\begin{abstract}

Sound scattering by a finite width beam on a single rigid body rotation vortex flow is detected by a
linear array of transducers (both smaller than a flow cell), and analyzed using a revised scattering
theory. Both the phase and amplitude of the scattered signal are obtained on 64 elements of the
detector array and used for the analysis of velocity and vorticity fields. Due to averaging on many
pulses the signal-to-noise ratio of the phases difference in the  scattered sound signal can be
amplified drastically, and the resolution of the method in the detection of circulation, vortex
radius, vorticity, and vortex location becomes comparable with that obtained earlier by time-reversal
mirror (TRM) method (P. Roux, J. de Rosny, M. Tanter, and M. Fink, {\sl Phys. Rev. Lett.} {\bf 79},
3170 (1997)). The revised scattering theory includes two crucial steps, which allow overcoming
limitations of the existing theories. First, the Huygens construction of a far field scattering signal
is carried out from a signal obtained at any intermediate plane. Second, a beam function that
describes a finite width beam is introduced, which allows using a theory developed for an infinite
width beam for the relation between a scattering amplitude and the vorticity structure function.
Structure functions of the velocity and vorticity fields deduced from the sound scattering signal are
compared with those obtained from simultaneous particle image velocimetry (PIV) measurements. Good
quantitative agreement is found.

\end{abstract}
 %\maketitle
%\begin{multicols}{2}
\narrowtext
\section{Introduction}

Ultrasound wave scattering provides a non-perturbative tool to probing a hydrodynamic velocity field.
Propagating sound waves are modified by their interaction with the flow that results in their
amplitude and phase distortions and scattering at different angles\cite{fil}. The sound scattering by
a flow was studied for the last 50 years theoretically\cite{lindsay,kraich,fabr,lund} as well as
experimentally\cite{grom,baudet,oljaca} with a goal to develop a reliable technique to directly study
dynamics of vorticity and velocity fields. Being developed and being reliable this tool can provide
rather unique possibility to get information about dynamics of turbulence. However, in spite of the
efforts the goal to measure the dynamical structure factor of the vorticity in a turbulent flow
remains rather far from reach. In this paper we attempt to address much less ambitious but rather
important experimental problem: is it visible to obtain reliable information about velocity and
vorticity field of an axisymmetric large vortex either stationary or time-dependent from a resulting
scattered signal of a finite width sound beam by a finite width receiver and rather
close to the scattering region?\\
 The full structure of the
scattering waves (amplitude and phase) can be related to a structure function of a vorticity via
analytical relation only in the Born approximation and in a far-field limit\cite{fabr,lund}. This
theory however considers a beam, a sound emitter, and a receiver of an infinite width. Only
Ref.\cite{berthet} discusses an influence of a finite width of a Gaussian sound beam on scattering
from a point vortex in a far-field limit. Finite width of these constituents particularly if they are
smaller than a flow region, leads to additional diffraction effects, which are mixed with interference
patterns resulted from the scattering and refraction. This unavoidable nuisance effects should be
unravelled from the intrinsic scattering signal, from which the vorticity could be determined. The aim
of the studies presented in the paper is to develop a kind of analysis, which can allow determining
the vorticity and velocity field without limitations described and to apply it to the experimental
data on the sound scattering from a single either stationary or time-dependent vortex.\\
 The majority of the experiments in the past\cite{baerg,ho,bass} were concerned with sound propagation
 in turbulent flows and
 determination of the scattering cross-section, which can be compared with theoretically predicted one.
  Such
 approach cannot provide unambiguous answer about the scattering by a turbulent velocity field. Later
 on it
 was realized that the scattering waves may be useful for remote probing of turbulent flows
 \cite{fabr,lund}. Using this
 idea the scattering problem was reformulated as the inverse problem to determine an instantaneous
 velocity field within a scattering region (laminar or turbulent) from measurements of the
 acoustic scattering at different angles similar to the light scattering problem. It was shown that in
 the approximation of a plane sound wave propagating through a velocity field with a continuous
 distribution of vorticity in a finite scattering domain  there is a linear relation
 between the Fourier component of the scattering sound wave amplitude and the
 spatial and temporal Fourier transform of the vorticity component normal to the plane of the wave
 propagation in a far-field limit\cite{fabr,lund}.\\
 During the years, several experiments were conducted to study either stationary or time dependent vortex
 flows by sound
scattering. Earlier experiments by Gromov et al\cite{grom} and later and more precise by Baudet et
al\cite{baudet} investigated the von Karman vortex street behind a cylinder at low Reynolds numbers,
$Re$. The experiments showed rather good qualitative agreement with the theoretical predictions. Later
on the experiment by Oljaca et al\cite{oljaca} on the ultrasound scattering from a stationary random
vortex flow formed along the axis of a swirling jet at $Re=10^4$, showed a good quantitative agreement
between measured and computed from separate velocity field measurements of the scattering  sound
amplitude. The main reason for the success was a satisfaction of rather strict limitations of the
theory, which leads to the linear analytical relation between Fourier transform of the vorticity and
the scattering acoustic signal\cite{fabr,lund}. In particular, it is required for an emitter to be
perfectly flat and wider than the flow region, and for a detector to be positioned in a far field
region. In the experiment\cite{oljaca} the characteristic size of the jet was about the sound
wavelength and much smaller than the receiver and the sound beam widths. Hence, the requirements were
perfectly met for the applicability of the theory. In this connection one can mention also
Refs\cite{dern,baudet1} where just structure functions of vorticity in a frequency domain were
measured. Another approach was undertaken in a series of papers by Fink and
collaborators\cite{fink1,fink2,fink3,fink4}, where the velocity field and vortex dynamics were
reconstructed from a phase shift measured by a transducer array. Time-reversal mirror (TRM) method is
used to amplify the effect of the flow on the sound phase distortion leading to a high resolution in
the velocity detection. The acoustic technique is based on a geometrical approximation, which is
justified for the flow field and sound characteristics used in the experiments. In a forward
scattering geometry the phase shift possesses the same information as a sound scattering but only in
the strictly geometric acoustics limit. Diffraction processes, which are relevant  for a finite wave
length as well as  finite width receiver and sound beam, require deeper understanding of sound
scattering particularly in a projection of an acoustic signal from a near-field to a far-field region.
On the other hand, there was a very recent attempt to achieve a goal similar to the formulated above,
by using direct numerical simulations to analyze the whole structure of the scattered wave (amplitude
and phase) and to compare it with the experimental measurements of the sound scattering on a vortex in
a turbulent flow\cite{fauve}. The agreement found is not great, particularly for the amplitudes. Our
approach is rather different: it is straightforward in acquisition of the
phase and amplitude and simple to apply especially at higher frequencies. \\

\section{Experimental set-up and measuring techniques}

The experiments were performed in a flow, which was produced inside a closed cylinder either with one
rotating disk or in the gap between two coaxial either co- or contra-rotating disks (the so-called von
Karman swirling flow). As detection techniques we use ultrasound scattering and particle image
velocimetry (PIV) to directly measure velocity field.\\

\subsection{Hydrodynamic set-up}

The hydrodynamic cell consists of a vertical Plexiglass (Perspex) cylinder of 29 cm inside diameter
and 32 cm in height (see Fig.1). A swirling water flow is produced between two rotating plates driven
independently by two motors with a maximum continuous torque of 13Nm. The motors are brushless
sinusoidal ones controlled by velocity mode drivers via optical encoders. Each shaft is centered and
is fixed by two bearings and sealed with Uniten rotating rings. The base of the device is suspended on
shock absorbers and the shafts are isolated from vibration by the motor couplings (Rotex GS designed
for servo systems). The acoustic components (emitters and a receiver array) are mounted flush to avoid
obstacles to the flow (see Fig.2). The cell is completely filled with deionized water with addition of
10ppm of surfactant. Small amount of surfactant facilitates removing of air bubbles from water.
Several steering configurations have been used: (i) Two steering plates of 100 mm diameter are
positioned 100 mm apart, each plate has four blades of 20 mm width and 5 mm thickness; (ii) the same
kind of plates with 40mm diameter and 4mm blades thickness; (iii) Upper plate is a metal disk of 280
mm diameter, a lower plate is absent, and only 25 mm diameter shaft is present with upper plane
located at 50 mm from the bottom of the cell; (iv) Two elongated shafts of 25 mm diameter are set 86
mm apart, grinded at the edge to a shape of one blade of 12 mm width and 6 mm thickness.

\subsection{Principle of measurements}

\subsubsection{Ultrasound transducers}

The measurement plane where the emitter and the detector array are placed, is chosen at the middle
plane between the disks. The sound detects the velocity field in the plane, or complimentarily,
 the vorticity field perpendicular to that plane. Ultrasound pulses are sent by a specially
designed emitter (see below) and detected by a linear array of 64 acoustic detectors with 1mm spacing
and 64x10mm overall active area (from Blatek). The acquisition system is built in a heterodyne scheme
meaning that 64 lock-in amplifiers are utilized on the incoming analog signals\cite{rinberg}. The
outgoing low frequency signals are passed to sample and hold components, which integrate the signal of
each pulse and hold the value to be scanned by two PC acquisition cards (National Instruments pci6071)
recording 128 channels, hence providing amplitude and phase components of the signal for 64 acoustic
channels. The lock-in amplifiers were compared to a SR844 RF lock-in amplifier (Stanford Research
Systems) and were found linear and reliable. The acquisition cards also function as controllers of
timing with precision of 0.05 $\mu$s. The detector and electronics are optimized for 5.5MHz. Each of
64 preamplifiers is based on VCA2612 chips from Texas Instruments, and it is built directly around the
detector. The entire circuitry consists of about 8,000 components. A special care was taken to block
electromagnetic interference emitted from the motors: (i) installing filters on the motors and the
power supply; (ii) the circuitry and the emitter cable are shielded in a triplex scheme where the
outer shell is connected to chassis ground, the intermediate shell connected to a one point analog
ground.\\
 We are able to measure amplitude and phase of the sound pressure simultaneously at 64 positions,
at various Doppler frequencies. The required transmitter activating signal is only 30mVrms (on certain
piezocrystals) , where the response of sound picked after
propagation and detection is around 500mVrms on the acquisition card.\\
 The time sequence of the
measurement process is implemented with two National Instruments pci6071 cards installed on a PC
running on Labview, and it goes as follows: A software key initiates once the sequence, in which a
rectangular pulse is sent repeatedly and activates a switch that connects the main signal to the
acoustic emitter. The pulse repetition rate is 505$\mu$s and its duration 18$\mu$s. The main signal is
sinusoidal with implied frequency between 220KHz and 6MHz. The emitted sinusoidal signal must remain
coherent between pulses and this will not be provided in burst mode of some function generators. The
rising edge of the pulse triggers additional three clocks (two event clocks are available in each
card).  1) A sample pulse is sent to engage the sample and hold components in the 64 lock-in
amplifiers. The pulse width is 8$\mu$s, which defines the integration time; delay time of 204$\mu$s is
chosen according to the acoustic propagation time with additional 5$\mu$s to allow for geometrical
delays, time shifts and sound from secondary sources to arrive on the detectors. The end of the sample
pulse triggers the acquisition scan by the cards. 2) A gain variation pulse is sent to the
preamplifier during the sample pulse and 1$\mu$s before in order to increase the gain to a preset
maximum value. Such step is taken to decrease the chance for a preamplifier to saturate and hang
during the sample time in response to sudden strong noise (e.g. reflection of side pointing beam from
the wall). 3) A clear pulse is sent to reset the sample and hold components by opening a path to
discharge their capacitors. The discharge starts after that sufficient time is allowed for the
acquisition cards to scan the channels (scan rate of 500KHz was used, and a waiting time was set
assuming the cards work in sequence). The discharge
time was set to 40$\mu$s (twenty times the characteristic discharge period).\\
 In order to avoid errors by
some arbitrary offset in the sampled signal a strobe mechanism was utilized. The frequency of the two
reference signals, which enter the product detectors in the lock-in amplifiers, was shifted by 100Hz
in relation to the main signal (coherence remains since the three function generators were connected
on the same 10MHz time-base). The sequence of pulses is not synchronized with any of the function
generators hence an initial arbitrary phase has to be resolved. In addition, since Labview on a PC is
not a real-time software it is not promised that the two cards begin acquisition in the same pulse.
The arbitrary phases are resolved by sacrificing two channels in each card. Thus, 62 lock-in
amplifiers are used for acoustic detectors and another two are connected to the main signal
(attenuated). During analysis the phase of the latter is subtracted from the former depending on which
card the lock-in amplifier is sampled. The direct hashing of the acquired data that provides the
amplitude and phase of sound is based on summing the signal around the peak in Fourier domain. The
center frequency is 100Hz as expected and the characteristic width is 50Hz. A summing window of about
100Hz is used, which is enough to include Doppler shifts by the time changing flow. Note that a zero
Doppler shift is expected for steady state flow (although locally the wavelength is changed by the
velocity of the flow). The given pulse repetition rate determines that only frequency shifts larger
than 1800Hz could alias with the main frequency
signal.\\
 Significant part of the efforts in building the probe were devoted to design a suitable acoustic
 emitter. The important parameters are the length of the transducer in the axis parallel to the
 measurement plane, and the smoothness of the phase and amplitude. A long transducer is necessary to
 produce
 a beam of parallel rays, to minimize the effect of side lobes, and to reduce the edge diffraction on
 the detector. The transducer lengths are varied between 3mm and 145mm. The transducer width is
 around 10mm, close to the width of the detector and determines the thickness of
 a layer, in which the flow is effectively averaged.\\
  A composite piezoelectric material was used in order to decrease lamb waves distortion. However,
 by using piezo-composite in the largest sizes a smooth phase still could be achieved but not the
 amplitude. The following types of transducers were built:\\
(T1) This type was used for the transducers of 3.5mm and 10mm lengths. A piezoelectric plate (SPC-Y02)
was covered by 4mm flat lens of Perspex and matched by conductive epoxy (Circuitworks) that was used
also
for wiring. The backing was air.\\
(T2) A piezoelectric plate (PZ34 from Ferroperm) was covered by a flat lens made of perspex with
groves for the soldered wires. The aperture size was 35mm, and the backing was air.\\
(T3) Two or three pieces of PZ34 plates were adjusted together side-by-side in a frame to make a
length of 96mm and 145mm, respectively. The backing was air, and the cover was a thin layer of epoxy.
Wires were soldered on the
edges.\\
(T4) A better transducer compared with T3 was built from polarized PVDF, of 125X10mm an active area,
mounted on Balsa wood and covered by a thin, conductive layer of epoxy and a thin layer of Acrylic.\\
 In all types of the emitters the enclosure was made of black Delrin, with 30dB/cm absorption.
 Types T1 and T4 were levelled with a round wall. The latter
 was made using a large opening made in the cell with house filled of water and covered by 10$\mu$m
 Mylar film.\\
Calibration of the array detector response in each acoustic element to a plane sound wave was
accomplished by moving a small emitter (19x6 $mm^2$ at 2.5 Mhz from Automation Industries) slowly and
precisely, paralleled to the array and taking measurements in steps of 1mm. Summation of the complex
wave function over all steps yielded the response to an ideal large emitter.\\

 \subsubsection{Particle Image Velocimetry (PIV)}

  The standard part of the PIV system was built by Oxford Lasers. A double exposure 1Mb CCD camera and
  a 15mJ
pulse infrared laser providing light sheet are used to measure the velocity field of the flow in two
dimensions, with particles as markers. The seeding particles are 20$\mu$m Polyamide from Dantec,
dissolved in a 50\% mixture with a surfactant (PolyOxyethylene-Sorbitan monolaurate from Sigma). The
plane of the velocity measurement is chosen to overlap the acoustic measurement plane. The analysis is
based on cross correlation between images by Fast Fourier transformation (FFT) of divided regions in
the frame (usually by 32X32 pixels). Analysis of each region provides one velocity vector, but the
algorithm can take overlapping regions up to 75\%, so that for regions of 32X32 pixels from a 1Mb
pixels camera a mesh of 128X128 vectors is obtained. In the implementation, the pixel equivalent size
was 80X210$\mu$m, the laser pulse duration was set to 60$\mu$s and the pulse separation was set to
2130$\mu$s. Thus, correlation between consecutive frames could fetch reliably velocities between
0.1-1.1m/s (in a worse case, using 32X32 pixels). There is a tradeoff between mesh resolution and
velocity accuracy and limits. Our analysis expands the velocity limit by combining results of analysis
with different area sizes of a correlation region. We limit our search to low velocity gradients and
thus we can interpolate higher velocities in a fine mesh that reveals the structure of the lower
velocities.\\
 The system was modified to use it in the narrow field of view available between the rotating plates.
Since the design of the experimental set-up (in two disks configuration) does not allow the camera to
look right above the light sheet plane, we were accounted with three problems. (i) The view angle
should minimize pickup  reflections from the wall. (ii) The converging effect of the circular wall of
the water filled cell as well as reduced optical clarity can be corrected by using windows made of
flat pockets of Perspex filled with water outside the cell. For reducing reflections separate windows
for the laser sheet and the camera were found preferable. (iii) Focusing on a plane in a perspective
view should use the Scheimpflug condition, in a way known as a perspective control. The perspective
control requires breaking the alignment between the camera and the lens so that the plane of the CCD
(image plane) meets the principle plane of the lens at the same level, where the object plane meets
with the other principle plane of the lens. The required tilting of the lens comes on the expense of
reduced light intensity entering the CCD camera. In order to increase the light intensity we used two
lenses. The tilted lens chosen was Mamiya RZ67 75mm/4.5 short barrel lens, capable for the perspective
control. On the digital camera, we fixed a micro Nikkor-Nikon 50mm/1:1.4 lens through a F-C mount,
which was shorten down to 11mm. This configuration allows us to focus on almost 15cm of the plane
 from 30cm distance, subject to setting obtuse viewing angle.\\

\section{Theoretical background of the data analysis}

\subsection{Sound scattering by a finite width beam.}
Following a derivation of an equation for sound scattering and refraction by a flow mostly due to
Kraichnan\cite{kraich} and Lighthill\cite{light} one gets in the first approximation of a small Mach
number, $M=v/c\ll 1$, the following equation for the sound scattering caused by the velocity field
alone:
$$\Delta\Psi+k_0^2\Psi=-2\rho_0\frac{\partial^2 (v'_iv_k)}{\partial x_i\partial x_k}.\eqno (1)$$

Here $\Psi$ is the complex wave function that represents the sound pressure oscillations generating at
frequency $\omega$ by the emitter, $k_0=\omega/c$ is the sound wave number, $\vec{v}$ and $\vec{v'}$
are the flow velocity and the velocity oscillations due to sound propagation, respectively, and
$\rho_0$ is the uniform fluid density. Eq.(1) was obtained in the approximation where low frequency
sound generated by the flow and the sound attenuation due to viscosity were disregarded. Further, we
assume 2D symmetry for simplicity, and due to the fact that the sound frequency is much larger than
any frequency in the flow a quasi-stationary case is considered. The influence of, e.g., moving
vortices in the flow shows up in a frequency shift (Doppler shift) relatively to the incident wave
frequency. The former usually is smaller or comparable with characteristic frequency of fluid flow and
is not considered here. In order to evaluate Eq.(1), we choose the axis $x_1$ along a local incident
wavefront direction, thus $\vec{v'}\parallel \hat{x_1}$ , and expand the source term. There is no
influence of the curvature in the wavefront direction on the derivatives within the first order in the
Mach number approximation\cite{ll}. Then taking into account that the wave length is the smallest
scale in the flow one gets for an incompressible flow the following equation (in a plane wave
approximation)

$$\Delta\Psi+k_0^2\Psi=[2k_0^2 {\vec{v}\cdot\hat{x}}-2ik_0\frac{\partial
(\vec{v}\cdot\hat{x})}{\partial x}]\Psi/c,\eqno(2)$$

where $\hat{x_1}$ is replaced by $\hat{x}$ and is used to define the local ray direction.\\
One can define the sound scattering field as $\Psi_{scat}=\Psi-\Psi_{rest}$, where $\Psi_{rest}$ is
the complex wave function in the absence of a flow that satisfies the following wave equation without
a source term:
$$\Delta\Psi_{rest}+k_0^2\Psi_{rest}=0.\eqno (3)$$

 By applying further the Green function method the solution of the scattering problem is
 obtained for a two-dimensional geometry in the following form (in $k_0|r-r'|\gg 1$ approximation) :
$$\Psi_{scat}=\frac{1}{c}\exp{(i\pi/4)}\int {d^2r'(k_0^2 v_x-ik_0\frac{\partial v_x}
{\partial x})\frac{\exp{(ik_0|\vec{r}-\vec{r'}|)}}{\sqrt{2\pi k_0|\vec{r}-\vec{r'}|}}\Psi_{rest}}.
\eqno(4)$$

In a far-field the solution can be drastically simplified and reduced to the analytical relation
between the scattering field and the Fourier transform either of the velocity or the vorticity fields.
Indeed, in the far-field $|\vec{r}-\vec{r'}|\approx |\vec{r}|-\hat{r}\cdot \vec{r'}$, and the integral
in Eq.(4) becomes the Fourier transform of the velocity and the velocity gradient fields. The variable
in Fourier domain is the scattering wave vector $\vec{k_s}=k_0(\hat{r}-\hat{x})$ and
$|\vec{k_s}|=2k_0\sin\theta/2$, where $\hat{r}$ is the unit vector from the center of the scattering
region toward the detector. The Fourier transform in 2D is defined as
$$F_{k_s}\{v_x\}\equiv \frac {1}{(2\pi)^2}\int {d^2r'\exp{(-ik_s\cdot r')}v_x(r')}.$$
Taking into account that $F_{k_s}\{\frac{\partial v_x}{\partial
x}\}=i(\vec{k_s}\cdot\hat{x})F_{k_s}\{v_x\}=ik_0(\cos{\theta}-1)F_{k_s}\{v_x\}$, one obtains in the
far-field limit

$$\Psi_{scat}=\frac{1}{c}\frac{(2\pi k_0)^2\exp{(i\pi/4)}}{\sqrt{2\pi k_0 r}}
\cos{\theta}F_{k_s}\{v_x\tilde\Psi_{rest}\},\eqno(5)$$

 where $\cos{\theta}=\hat{r}\cdot\hat{x}$ and $\tilde\Psi_{rest}=\Psi_{rest}\exp{(-ikx')}$. We
 present in eq.(5) the scattering field as a convolution of the Fourier transforms of the velocity
 field and the known beam function, $F_{k_s}\{\tilde\Psi_{rest}\}$. Only in an ideal case of an infinite
 planar wave front, when $\tilde{\Psi}_{rest}=1$, there exists a simple relation between the Fourier
 transforms of the velocity and the vorticity fields:
 $$F_{k_s}\{v_x\}=\frac{i(\hat{y}\cdot\vec{k_s})}{k_s^2}F_{k_s}\{(\nabla\times
 \vec{v})_z\}=\frac{i}{2k_0
 }\cot{(\theta/2)}F_{k_s}\{(\nabla\times\vec{v})_z\},\eqno(6)$$
which leads to a direct linear relation between the sound scattering field and the Fourier transform
of the vorticity field as first pointed out by Lund and Rojas\cite{lund}.

This general approach can be applied in a case of a single rigid body rotation vortex with a core
azimuthal velocity distribution as
$$v_t(r)=\Omega r, \hspace{1cm} {\mit for}\hspace{1cm}
 r\leq r_0, \eqno(7)$$

and outside the core up to the cell wall the azimuthal velocity component decays as

$$v_t(r)=\frac{\Omega r_0^2}{r_c^2-r_0^2}(-r+r_c^2/r),\hspace{1cm}{\mit for}\hspace{1cm} r_0<r<r_c,
\eqno(8)$$

where $\Omega$ is the angular speed, $r_c$ and $r_0$ are the cell and the core radii, respectively.\\
In the axisymmetric case one gets the following result for the Fourier transform of the $x$ velocity
component, $v_x$:
$$F_{k_s}\{v_x\}=\frac{-i}{2\pi}\cos{(\theta/2)}\Omega\frac{2r_0r_c}{k_s^2(r_c^2-r_0^2)}
[J_1(k_s r_0)r_c-J_1(k_s r_c)r_0], \eqno(9)$$ where $J_1(x)$ is the Bessel function, which was
introduced via $F_{k_s}\{v_x\}=\frac{-i(\hat{y}\cdot\vec{k_s})}{2\pi k_s}\int{r'v_t(r')J_1(k_s
r')dr'}.$  The expression (9) exhibits even for the unbounded beam width and detector length two
maxima in the scattering signal at some value of $k_s$ due to a final size of the flow region, $r_c$.
It is clear from  Eq.(9) that in a general case the location of the maxima
 depends on two parameters, $r_0$ and $r_c$. So in the case when $r_0/r_c\ll 1$, the peaks are located at
 $k_sr_c\simeq 1.11\pi$. This value alters by change of $r_0$. So the numerical calculations based on Eq.(9)
 give the following results (for $r_c=145$ mm): at $r_0=6$mm the peak position is at $k_sr_c\simeq 1.11\pi$,
 at $r_0=20$mm  the value is $k_sr_c\simeq 1.09\pi$, at $r_0=50$mm  the expected value is
 $k_sr_c\simeq 1.02\pi$, but at $r_0=100$mm the location value is already shifted to $k_sr_c\simeq 0.87\pi$.\\
The peak height, $H$, also provides useful information about the flow. Indeed, in the case of
$r_0/r_c\ll 1$ and of the infinite beam width, $H$ is proportional to $\Omega$, to a core circulation,
 $\Gamma=2\pi\Omega r_0^2$, and independent of
$k_0$, or sound frequency. The dependence on $\Gamma$ and independence of $k_0$ are rather non-trivial
results taking into account rather complicated functional dependence of $F_{k_s}\{v_x\}$ on the
parameters $r_0$, $r_c$,
and $k_0$.\\
  In the limit $r_c\rightarrow \infty$ the peak locations approaches zero angle
asymptotically but the value of the amplitude of the Fourier transform of the velocity at $\theta=0$
remains zero. In the limit of $k_sr_c\ll 1$ and at $r_0\ll r_c$ one can derive an asymptotic
expression for the Bessel functions in Eq.(9) and get $F_{k_s}\{v_x)\}|_{k_sr_c\rightarrow
0}=-\frac{i\Omega}{8\pi}r_0^2r_c^2k_0\sin\theta/2$. From this expression it is easy to see that the
amplitude at small angles increases proportionally to $\Omega$, $\Gamma$, and $k_0$, and approaches
zero at $\theta\rightarrow 0$. \\
 For a finite width beam (or emitter) or a finite width detector the Fourier transform of the scattering field
 wave function from Eq.(5) can be expressed as a convolution $\Psi_{scat}(\theta)\sim
 F_{k_s}\{v_x\}\ast F_{k_s}\{\tilde\Psi_{rest}\}$. In a case of the finite width, $d$, (either of a beam
 or a detector whatever is the smallest) let us consider first the Gaussian beam model. The beam function
 is defined as $ \tilde{\Psi}_{rest}=\exp{(-y^2/[2(d/2)^2])}$. For $d/r_c\gg 1$
the behaviour of the peak height here is similar to the infinite width beam considered above.\\
 The location of the peaks in a finite
width Gaussian beam depends on the beam width: at $d=50$mm the peak position is at $k_s d\simeq
0.94\pi$ (at $r_0/d\ll 1$). When $r_0$ increases this relation is changed: at $r_0=10$mm the peaks are
located at
 $k_s d\simeq 0.92\pi$, and at $r_0=50$mm they are found at $k_s d\simeq 0.71\pi$. The dependence of
 the peak heights on the parameters for a finite width beam is more elaborate, since here three
 characteristic lengths exist in the problem: $r_0$, $r_c$, and $d$. In spite of this fact the approximate
 scaling law was found from numerical calculations. It follows that $H$ is independent of $k_0$ and
 proportional to $\Gamma$ with the scaling function $g(x)$ (where $x=r_0/d$) that is different for two
 regions of small and large beam width:
  (a) $g(x)=125x^{1.2}$ for $x>1$, and (b) $g(x)=66+66x^2$ for $x<1$.\\
In the case of a finite rectangular beam (or a finite detector whatever is the smallest) the peak
location is defined by $k_sd$ as in a finite Gaussian beam considered above. At $r_0/d\ll 1$ one gets
$k_s d\simeq 1.52\pi$ (e.g. for $d\simeq 50$mm). For larger values of $r_0$ the peak location is
shifted towards smaller values. So, at $r_0=10$mm the estimated peak position is at $k_s d\simeq
1.48\pi$, and at $r_0=20$ and 50 mm one gets $1.39\pi$ and $1.34\pi$, respectively. The peak height
similar to the case of the Gaussian finite width beam, is independent of $k_0$ and proportional to
$\Gamma$ with the scaling function $g(x)$ that is slightly different from the former case. It is
determined for the range of arguments values $0.01<x<4$ at $d>30$mm and $r_c=145$mm and has the
following form: (a) $g(x)=205 x^{1.2}$ for $x>0.5$, and (b) $g(x)=63+103x^2$ for $x<0.5$.\\
In order to demonstrate the influence of a finite beam width on the scattering signal from a  single
vortex with a core radius $r_0$, we present in Fig.3 the results of numerical simulations for the
structure functions of the velocity field for the vortex of $r_0=5$mm and three beam widths of 20, 60,
and 120 mm (but all of them smaller than the cell size, $r_c=145$mm) compared with an infinite beam
width for the same cell size. Only for $d=120$ mm the wave number of the structure function peak
location becomes close to the peak location for the infinite width beam (see Fig.3a). The wave number
of the peak location in the infinite width beam case is defined by the cell size (see explanation
above after Eq.(9)). One can see that in spite of the fact that $d\gg r_0$, the finite width of the
beam drastically alters the resulting scattering signal. Thus in order to get correct results on the
structure functions of the velocity and vorticity fields one should use either a beam width exceeding
the flow size, $r_c$, or to perform a deconvolution based on a provided constrain, which fills out the
missing information about the velocity field outside the beam extent.

 \subsection{Alternative approach to find velocity and circulation: phase shift slope
  in a forward scattering.}

While the study of scattering data requires to extract from  sound
signals both the amplitude and the phase variations, one can find
information about the velocity field and the circulation from the
phase shift. We consider two cases in a phase shift analysis. In
the first case of a large emitter, rays remain parallel in
propagation, and the phase shift as function of $y $ (assuming it
can be measured avoiding interferences between the paths) can be
defined in the limit $M\ll 1$ as\cite{ll}:
$$\phi(y)=(k_0/c)\int{v_x(x',y)dx'}.\eqno(10)$$

 Lindsay has found\cite{lindsay} that
the phase shift induced by a point vortex has a linear dependence on an angle. This result can be
obtained if we consider rays passing outside the core, and assume a large emitter at $x\rightarrow -
\infty$ and velocity profile $1/r$, namely:
       $$v_x=\frac{\Omega r_0^2 y}{x^2+y^2}.\eqno(11)$$
Phase shift is measured on a screen at $x=r_c$, with  $\theta=\tan^{-1}(\frac{y}{r_c})$, then the
slope can be found, outside the singularity at $\theta=0$, with the following constant value:
     $$\frac{\partial \phi}{\partial
     \theta}=\frac{\partial}{\partial\theta}\int_{-\infty}^{r_c}{\frac{k_0 v_x}{c}dx}=-\frac{k_0}{c}\Omega
     r_0^2.\eqno(12)$$
Our idea is to look in particularly on rays passing near the center of the vortex at $\theta=0$ and
try to get information on the flow from the phase shift slope. For an infinite source at $x=-r_c$  and
detectors at $x=r_c$:
     $$\frac{\partial\phi}{\partial\theta}
     |_{\theta\rightarrow 0}=\frac{k_0
     r_c}{c}\int_{-r_c}^{r_c}{\frac{\partial v_x}{\partial y}dx} \hspace {1cm} {\mit at}\hspace{1cm}
     y\rightarrow 0.\eqno(13)$$
Thus, substituting the velocity field projected on $x$ via Eqs.(7,8), one obtains the phase shift
slope as
$$\frac{\partial\phi}{\partial\theta}|_{\theta\rightarrow 0}=\frac{k_0 r_c}{c}4\Omega
r_0(1-\frac{r_0}{r_c+r_0}).\eqno(14)$$

 Comparing Eq. (14) and Eq.(12) we find that a rigid body rotation core induces a larger slope
 by an order $r_c/r_0$ compared with the exterior of the vortex. Thus for a
point vortex at $\theta=0$  there is a step in the phase shift, having an infinite slope, revealed
therefore as the Berry's phase\cite{berry}.

A total change in the phase shift between two points on the acoustic screen can be calculated by a
loop integral over the velocity field. Since a velocity along the detector and the emitter is zero,
there is a simple relation between the circulation and the step in the phase shift followed from Eq.
(10)\cite{sonin}:
$$\Gamma=\oint{v\cdot dl}=\frac{c}{k_0}(\Delta\phi_{(2)}-\Delta\phi_{(1)}),\eqno(15)$$
where $\Delta\phi_{(1)}$ and $\Delta\phi_{(2)}$ are the phase shift differences on two sides of the
detector, respectively.\\
 In the second case we consider rays diverging out of a single point on a
small emitter (and there is an axisymmetric single vortex). We can show that the phase shift slope,
calculated from the phase shift due to a vortex at the cell center between a detector placed at
$x=r_c$, and sound emitter located at $(x,y)=(-r_c,0)$, can be written as
                           $$\frac{\partial\phi}{\partial\theta}|_{\theta\rightarrow
                           0}=\frac{k_0 r_c}{c}\int_0^{r_c}{\frac{v_t(r)}{r}dr}.\eqno(16)$$
The angle $\theta$ is measured between the radius from the center of the vortex and the beam
direction. The phase difference used to derive Eq. (16) can be written as\cite{lindsay}
$$\Delta\phi(\theta)=k_0\int_{-r_c}^{r_c}{(n-1)}ds',$$
where $n=\sqrt{1-2\hat{s}\vec{v}/c}$  is the refraction index in the first approximation in the Mach
number, $\hat{s}$ is the unit vector in the ray direction, and $s'$ is the coordinate in the ray
direction. Then from simple geometrical considerations one gets Eq. (16). Substituting the expression
 for the velocity profile Eqs.(7,8) into Eq. (16) we obtain:
               $$\frac{\partial\phi}{\partial\theta}|_{\theta\rightarrow
                           0}=\frac{k_0 r_c}{c}2\Omega r_0(1-\frac{r_0}{r_c+r_0})\eqno(17)$$

that is equal exactly a half of the result for a large emitter.

\subsection{Construction of a far field wave function from the near field measurements.}

 Probably, the most severe limitation of the theory to be applicable to an experiment is
the far-field approximation. In this respect the relevant question is how to extrapolate sound
scattering measurements obtained at a distance not far away from  a scattering region into a far-field
region to be compared with the theory and to reliably extract information about the vorticity
structure function. Our validation of the far-field scattering result is based on comparison with
calculations, which use the velocity field measured by PIV technique.

 A far-field construction of the scattering field from the acoustic field (either pressure
or scattering wave function) given at a certain plane in 3D case and at a certain line in 2D case, is
based on the mathematical description of a Huygens principle typically used to describe radiation from
a curved surface. This problem is similar to the problem of  diffraction by a thin screen of finite
dimensions (or for a finite dimension wave beam)\cite{mar}. Thus our goal is to consider the
propagation of a wave into unbounded source-free half-space $x\geq 0$, when certain conditions in the
initial plane $x=0$ are specified. Then the scattering field in 2D at the location $x\geq 0$ can be
defined from the following Rayleigh-Sommerfeld integral\cite{mar}:
$$\Psi(x,y)=- \int_{\Sigma_+, \Sigma_-}{\Psi(0,y')\frac{\partial}{\partial
x}[G(x,y;0,y')]}dy',\eqno(18)$$

where $G(x,y;x',y')\equiv G(|\vec{r}-\vec{r'}|)=\frac{i}{4}H^{(1)}_0(k|\vec{r}-\vec{r'}|)$ is the
Green function of the Helmholtz equation in two dimensions(2D), $\vec{r}=(x,y)$, and $H^{(1)}_0 $ is
the Hankel function of zero order. Equation(18) follows from the integral Helmholtz theorem, where use
of the Green's function is made. $\Sigma_+$ and $\Sigma_-$ denote the two sides of the diffraction
screen ($x>0$) and ($x<0$), respectively. The presence or absence of the screen is not essential to
the derivation of eq.(18) but in the absence of the screen one does not need to apply the Kirchhoff
approximation. To get a more simplified expression in a far-field region the asymptotic expression for
the Hankel function can be used $H^{(1)}_0(z)_{z\gg 1}\approx \sqrt{\frac{2}{\pi z}}\exp{i(z-\pi/4)}$.
Then the far-field construction of the scattered sound wave function can be calculated in the limit of
$k_0|\vec{r}-\vec{r'}|\gg 1$ and $k_0(y-y')^2/r\ll 1$ as
$$\Psi(x,y)_{ff-scat}=k_0\sqrt{i}\int{dy'\frac{i}{\sqrt{2\pi
k_0(r_f-r_d)}}\exp{(\frac{ik_0(y-y')^2}{2(r_f-r_d)})}\Psi(0,y')_{d-scat}},\eqno(19)$$

where $r_d$ and $r_f$ are the distances measured from the cell center till the detector and the
far-field region, respectively, $\Psi(0,y')_{d-scat}$ and $\Psi(x,y)_{ff-scat}$ are the scattering
wave functions at the detector (as measured) and at the far-field, respectively. Since at the array
detector there are just 64 elements, a
sum instead of the integral is used to calculate the wave function in the far-field.\\
 The method presented is found advantageous over other inverse solution methods. In
 measurements as well as in numerical simulations it is also desirable to apply tapering of the beam edge
at the emitter to avoid an oscillating edge diffraction pattern.\\

\section{Experimental results.}

\subsection{Method of data analysis}

In order to get information about the scattering sound amplitude one needs to find a phase difference
between an incident and a scattered signals. Our way to find the scattering wave is simply to measure
and subtract the complex sound signals with a flow and at a rest. At high enough frequencies we can
concern ourself only with the main change in the wave, which is the phase shift due to change in time
of flight by the flow. The phase shift is found from the phase information as
$\phi=-\arg(\frac{\Psi_{flow}}{\Psi_{rest}})$.

The amplitude of scattering wave on the array detector is determined as $A=|\Psi_{flow}-\Psi_{rest}|$.
Both wave functions, $\Psi_{flow}$ and $\Psi_{rest}$ are the central frequency component of a sound
extracted from a bank of pulses entered to a Fourier type filter, as described above. Information
obtained during a period when the bank of pulses is collected, is called a frame. In a steady flow we
average the complex results between available frames. For the dynamic study of a flow we compare the
wave function in a flow with the average on many frames to avoid suspecting additional sources of
fluctuations. Another step before substitution of $\Psi_{rest}$  into the last equations is to
normalize it by a factor $\exp{i\gamma}$. The phase correction $\gamma$  is the average in time
(between frames) of the values $\gamma'$ that minimize the expression
$<|\Psi_{flow}-\Psi_{rest}\exp{i\gamma'}|>_y$ (average on y refers to channels of the array).
 Such step is used to specifically regularize results of scattering from a single vortex and it also
 compensates for occasionally uncontrolled conditions that
change the overall refraction (for example raising of seeding particles by the flow, when they are
used). At the end, we obtain the scattering wave function on the transducer array plane as:
$$\Psi_{scat}=\Psi_{flow}-\Psi_{rest}.\eqno(20)$$
The presented results of scattered and incident fields are normalized so that the incident field on
the emitter plane is of a unit amplitude. To calculate the incident field at a distance $l$ from its
plane we use the Huygens construction (see Eq.(18)) that is a propagation transform of the window
function (with 5\% apodization at the edges to reduce signal corruption due to noise) $\Psi(0,y)$:
$$\Psi_{rest}(l,y)=ik_0\sqrt{i}\int {dy'\exp({ik_0\sqrt{l^2+(y-y')^2}})\frac{1}{\sqrt{2\pi
k_0l}}\Psi_{rest}(0,y')}.\eqno(21)$$ Then $\Psi_{scat}$ can be calculated from Eq.(4).\\

\subsection{Characterization of the flow by PIV.}

We focus on a single vortex, created between co-rotating plates, and find its average properties in
time. Since the vortex is axisymmetric we can extract from PIV analysis the profile of an azimuthal
velocity versus the radius. The center of the vortex is found according to a minimum in the orthogonal
projection of a velocity vector on x and y axes (except for the smallest vortices produced by 25mm
rods, where we designated the center by eye recognition). The profile is built by averaging on 64 PIV
maps, such as shown in Fig.4a. The profiles for the three inspected flows are shown in Fig.4b
(co-rotating 25mm rods), Fig.5 (co-rotating 40mm plates), and Fig.6 (co-rotating 100mm plates). In
general, we see that rotating plates with blades produce a core of a rigid body rotation, i.e., the
core velocity increases linearly with the radius. Outside the core, the velocity decreases like in a
flow between two cylinders.  This solution (Eqs. (7,8)) was used to fit the measured velocity profile,
and the flow parameters were extracted from
 the parameters of the fit. Thus we find from the fits $r_0=6$mm, $r_0=14$mm and $r_0=42$mm, respectively.
 In all cases generally the core rotation frequency is the same as of the motor.
The PIV maps extend to about third of the cell due to geometrical limitations in a field of vision and
a limited power of the laser (nominally 15mJ/pulse). Therefore, the velocity profile is extrapolated
using the model of a flow between two cylinders. The distance from the cell center to the detector's
plane and the wall is $r_{c}=145$mm. We used an average on 64 velocity maps of the flow for 40mm
plates (extrapolated between radii of 50mm and 145mm) to evaluate the integral of the orthogonal
velocity projection on x-axis. The result can be compared to the ultrasound phase shift measured at
5.5MHz, as shown in Fig.7.

\subsection{Velocity and circulation measurements via phase shift.}
  As follows from Eq.(15) the measurement of the phase shift provides direct information about the
  circulation. This result can be obtained from the data in Fig.7, where the spatial dependence
  of the phase shift obtained from the sound and PIV (here the velocity field measurements
  were converted into the phase shift via Eq. (10)) measurements are presented. We would like to
  emphasize here that all experiments presented and analyzed were realized in the limit $M\ll 1$,
  so that the expression for the phase shift delivered in Eq.(10) is applicable. It follows that
  for the maximum
  phase difference measured in Fig.7 of about $\Delta\phi=0.7$rad, the maximum circulation is about
  $\Gamma=45,000mm^2/sec$. As one can find from Fig.7 the value of the maximum phase difference and
   correspondingly the maximum circulation coincide with those found from the PIV measurements
   within 1-2\%.\\
  The phase shift patterns due to the flow based on averaging of 10 frames of 512 pulses for various
 emitter sizes were examined. It appears remarkably for two types of flow considered (both a single
  vortex flow with a rigid core rotation, produced by 100mm plates and 25mm rods) that the
  slope of the phase shift around the center of
  the beam versus scattering angle (or a distance along the detector) is constant. We describe in
  detail the flow produced between two co-rotating 100mm plates, which provides stable phase
shift plots. In Fig.8 we demonstrate how a value of the phase shift slope
$\partial\phi/\partial\theta$
 are obtained: the limit is chosen to give a constant slope with the smallest linear regression error.
  According to Eq.(17) the phase shift slope provides direct information about the vorticity of the
 vortex, $\tilde{\omega}=2\Omega$. The results for emitters of different sizes and at various
 frequencies are shown in Figs.9a-d. The
  phase shift slopes for all frequencies and all emitters appear to be linearly dependent on the
  rotation speed, $\Omega$. The data presented can be either scaled by $\Omega$ or presented
  via the derivative
  on $\Omega$, $m\equiv\frac{\partial(\partial\phi/\partial\theta)_{\theta=0}}{\partial\Omega}$. Then
  the data on the proportionality coefficient $m$ as a function of frequency, calculated from the
  plots presented in Figs.9a-d, are
  summarized in Fig.10. The proportionality coefficient, $m$, depends linearly on the frequency in
  the range between 0.5 and 5.5MHz, and the results are sharply separated in two groups:
  (i) small emitters (3.5, 10mm), and (ii) large emitters (35, 145mm) compared with the size of
  the detector array
  of $L=62$mm. The rotation speed of the plates is used to define the flow parameter and is
  converted to the angular speed of the vortex core obtained from the fits of the PIV data.
  According to Eq.(17) in
  the theoretical section, we expect a point-like emitter to have the frequency dependence
  of the proportionality
  coefficient, $K=\partial m/\partial f$, on the plots of Fig.10 as follows:
$$K\equiv\frac{\partial^2(\partial\phi/\partial\theta)_{\theta=0}}{\partial\Omega
\partial f}=\frac{4\pi}{c^2}\frac{r_{c}^2r_0}{r_c+r_0}=0.0281(sec)(MHz)^{-1}.$$

 The experimental value of $K$ for 3.5mm and 10mm emitters is $K=0.026\pm 0.001$
 sec(MHz)$^{-1}$. The value of $K$ from the plot in Fig.10 for 35mm and 145mm
 emitters is $0.045\pm 0.002$ sec(MHz)$^{-1}$. As explained in section IIIB the theoretically
 predicted value of $K$ for a small emitter should be twice smaller than for a large one, i.e.
 the theoretically expected value for the large emitters is $K=0.056$sec(MHz)$^{-1}$ according
 to Eq. (14). However, specifically for these
 measurements the large emitters blocked 12\% of the total diameter of the exterior of the flow.
 Assuming effective reduction in the cell radius $r_c$ the expected theoretical value should be
  corrected down to $K=0.048$ sec(MHz)$^{-1}$, and it becomes rather well comparable with the experimental
  value presented above.

\subsection{Sound scattering data.}

Using Eq.(4) derived in the theoretical section we were able to calculate the scattered signal
amplitude from a given velocity field assuming that the emitter wavefront is known exactly. We used a
velocity profile of a flow produced by 100mm rotating plates extracted from PIV (Fig.6).  The emitter
of 35mm length was used and modelled for a flat amplitude (a plane wave) as the input, to obtain the
result in Fig.11a. The measured incident and scattered wave amplitudes at the frequency 4.0MHz are
compared to the calculated ones (they are scaled to unity value of the wave amplitude at the emitter
exit)(see Fig.11a,b). In the calculation of the sound scattering field, a mesh of a half wavelength
resolution was used in a zone limited to twice of the emitter size and the velocity field was
interpolated based on the velocity profiles values. The wavefront was calculated at various
cross-sections in the cell using Eq.(21). The measurement was performed at 5.5Mhz collecting $20\times
512$ pulses in a flow produced by the 100 mm disk at the rotation speed $\Omega=22$ rad/sec. In
Fig.12a,b similar data of the incident and scattering sound fields from 125mm emitter at the frequency
5.5MHz are shown (100 mm disks, $\Omega=35$ rad/sec). As in the previous case an agreement with the
theoretical calculations is rather good. An evidence of a diffraction pattern (side-lobe) from the
emitter of 3.5mm length at the frequency of 5.5MHz in still water (regarded as the incident wave), and
in the scattering signal due to vortex flow between the rods of 25mm diameter at $\Omega=71$rad/sec,
is clearly seen in Fig.13a,b. The direction of the initial sound beam was slightly tilted relatively
to
 the direction to the cell center, so that the center of the beam is shifted from the center of the
 flow. This fact was not regarded in the fitted curve that results in a discrepancy of the measured
 and calculated sound amplitudes. This result clearly demonstrates that the scattering signal can be
 easily buried in the incident signal coming from the side-lobes at sufficiently large angles of
 detection.

\subsection{Far field construction of sound scattering field and its relation to velocity and vorticity
 of the scattering region.}

We use the Huygens projection of the near-field scattering signal detected at the receiver array
plane, into a synthetic far-field plane using the Rayleigh-Sommerfeld integral via Eq.(19). We choose
the far-field plane at a distance of $r_{ff}=250$m, much larger than $d^2 k_0/8\pi=3$m. The signal is
tapered 2mm on each edge before the projection to suppress numerical instability of the diffraction
pattern. By this procedure one obtains full information of an amplitude and a phase of the scattered
signal in the far-field. However, practically due to finite spatial resolution the phase information
becomes corrupted. So the next step in the far-field construction is to replace the phase variation as
a function of an angle (or a wave number) by a function toggled at each minimum point by a $\pm
\pi/2$. In such way we reconstructed the phase field using an observation made in the simulations that
at every minimum point of the correct amplitude curve the sign of the field should be inverted. In
order to avoid some spurious minimum points the number of digitizing points in integration was
increased, and the result was compared
against finer digitization mesh.\\
 According to Eq.(5) the sound scattering field in the far-field is proportional to the
Fourier transform  $F_{k_s}\{v_x\tilde\Psi_{rest}\}$, i.e. in such a way plots of the modulus of the
two-dimensional Fourier transform,  $|F_{k_s}\{v_x\tilde\Psi_{rest}\}|$, in the entire cell  were
obtained in Figs.14,15. The data projected into the far-field are compared with calculations of the
scattering field based on the velocity field measurements by PIV. The results are presented in a view
angle observed from the center of the cell through the receiver as an aperture. This angle, $\theta$,
is related to the scattering wave vector, $k_s$, via formulas: $k_{sx}=k_0(\cos\theta-1)$,
$k_{sy}=k_0\sin\theta$, and $k_s=2k_0\sin\theta/2$.\\
We studied properties of the modulus of the Fourier transform,  $|F_{k_s}\{v_x\tilde{\Psi}_{rest}\}|$,
at different sound frequencies and different rotation speeds. It was revealed that the angular
location of the main peaks in $|F_{k_s}\{v_x\tilde{\Psi}_{rest}\}|$ is proportional to the sound
wavelength (Fig.16). The experimental value of the slope of the plot is $2\sin{(\theta_p/2)}(k_0
d)\equiv k_s d=(1.51\pm 0.11)\cdot\pi$, and is found to be in a good agreement with
 the value of the slope obtained from numerical simulations $k_s d=1.52\cdot\pi$ (see discussion
 in Sec.IIIA). It was also found that
the scaled value of the peak height $P=Hd^{-0.8}g(r_0/d)(145/r_c)^{0.2}$ of
$|F_{k_s}\{v_x\tilde{\Psi}_{rest}\}|$ is proportional to the circulation $\Gamma$ (Fig.17). Thus, four
different sets of the data for three different vortices and two beam sizes can be scaled down using
the functional dependencies on the beam width, $d^{0.8}$, and the reduced vortex core size,
$g(r_0/d)$, taken from our numerical calculations (see discussion in Sec.IIIA).  As seen from the plot
scaling does not work great for one set of the data, possibly, due to flow imperfection. On the other
hand, the proportionality of the scaled peak height to $\Gamma$ is not so obvious as one can decide
from the first sight, since the finite beam width described by the beam function
$\tilde{\Psi}_{rest}$, alters significantly the function presented in Eq.(9). We also found from the
measurements that the peak height, $H$, is independent of the sound frequency and depends strongly
(about $d^2$ when $d<2r_0$) on the beam width. When $d\gg 2r_0$ the peak height value, $H$, is
proportional to $r_0^2$ in a full agreement with the simulations. It is obvious that the finite size
of either a beam (an emitter) or a detector (whatever the smallest) limits our knowledge about the
scattering field and, therefore, also about the Fourier transform of the velocity field at the wave
numbers (or angles) smaller than that
corresponding to the peak location..\\
 The projected scattering to a far-field, $F_{k_s}\{v_x\tilde{\Psi}_{rest}\}$, does not have a direct
 relation to the structure
function of a vorticity, $F_{k_s}\{(\nabla\times\vec{v})_z\}$.  In order to get the latter, one should
extract information about a 2D Fourier transform of the velocity field $v_x$ from the measurements of
$F_{k_s}\{v_x\tilde{\Psi}_{rest}\}$.  However, this transformation is a singular one, so a choice of
constrains is made roughly related to the continuity and irrotaional character of the flow outside of
the acoustic beam extent, where $\tilde{\Psi}_{rest}$ decays to zero. Particularly for the plane beam,
a 1D Fourier transformation is relevant, since in the propagation direction, $x$,
$F_{k_s}\{\tilde{\Psi}_{rest}\}$ can be presented by a delta-function in the wave number domain. We
performed a backward and forward 1D Fourier transformation of $F_{k_s}\{v_x\tilde{\Psi}_{rest}\}$
values on a set of points in $\theta$ coordinate (which span about $90 \deg $), such that the values
are not modified initially. Next we added the filtering effect that removes $\tilde{\Psi}_{rest}$ by
extrapolating the intermediate result
$\hat{v}_x(y)=2\pi\int{dk_{sy}\exp{(ik_{sy}y)}F\{v_x\tilde{\Psi}_{rest}\}}$ as if it were
one-dimensional projection of the velocity,  $v_x^{1D}(y)$, using our flow model for extrapolation
(see Fig.18). The phase in the extrapolated region of $\hat{v}_x(y)$ should be adjusted having
continuity at $y=\pm d/2$ and a $\pi$ phase difference at the tails. Such construction provides the
required velocity field and correspondingly its Fourier transform in the wave number domain,
$F\{v_x\}=\frac{1}{(2\pi)^2}\int{dy\exp{(-ik_{sy}y)}\hat{v}_x(y)}$. Now using Eq.(6) one can relate in
such a way the obtained $F_{k_s}\{v_x\}$ to the required structure function of the vorticity
$F_{k_s}\{(\nabla\times\vec{v})_z\}$ (see Fig.19a). With a help of the Wiener-Khinchin theorem one can
also obtain an azimuthally averaged point by point correlation function of the vorticity, which in the
case of an isotropic flow or axial symmetry gives
$$C_v(r)=\frac{k_0^2}{\pi}\int{dk_sJ_0(k_sr)|F\{v_x\}|^2\tan^2{(\theta/2)}}. \eqno(22)$$
 The corresponding correlation function of the vorticity is shown in Fig.19b. Since the weight of
errors in estimation of $F\{v_x\}$ is more pronounce at high scattering angles, the integral in
Eq.(22) should be cut in a tight range. We used a Gaussian attenuation factor (with $k_s=0.7 mm^{-1}$
characteristic value) on the far-field scattering values to filter out high
frequency fluctuations.\\
  It is obvious that the vorticity structure function restored in such a way
contained additional information reflecting our guess (or our PIV results). However, this additional
information is mostly relevant to a low wave number range of the vorticity structure function, i.e. to
the core of the Fourier transform of the vorticity. Therefore, at wave numbers larger than the core
wave number value, the vorticity structure function provides information based on the original
experimental data.

\subsection{Dynamics of a vortex measured by sound scattering.}

The sound scattering technique can be used to study vortex dynamics. It can be studied by a phase
dynamics approach to get temporal variation of a vortex location, vortex radius, and vortex
circulation similar to what was done in Ref.\cite{fink2,fink3}. However, we used a different approach
of sound scattering and compare the results on a vortex precession with those obtained by PIV. The
vortex position is found from the minimum point of the scattering amplitude pattern. The periodicity
in the vortex motion is characterized by the variation of the peak heights in the far-field scattering
pattern. The setup is used with the upper plate of 280mm in diameter at the angular speed
$\Omega=30$rad/sec, and the lower plate is absent. The rate of PIV tracking is one map per 0.266
seconds. The rate of ultrasound tracking based on collection of 32 pulses per frame is one wave
function plot per 0.016 seconds (sound frequency is 2.5MHz, emitter is 96mm long). In the case of the
rotating upper plate, a periodic precession of the vortex is found by both techniques: PIV and the
sound scattering (see Fig. 20). To compare two sets of the data the time correlation functions of the
vortex locations were produced from both sets. A good quantitative agreement between the dynamic
results of PIV and ultrasound measurements was found (see Fig.21)with a period of 0.90sec. Besides,
the peak height of the Fourier transform of the velocity projected into a far-field is found to be
twice periodic compared with the vortex position (see Fig.22a). It is clearly seen in the
auto-correlation function presentation of the data in Fig.22b. This effect occurs due to increase or
decrease in the integral of $v_x$ over the beam area (detector view), when the vortex core is shifted
sideways, since for small $k_s$ $|F_{k_s}\{v_x\tilde{\Psi}_{rest}\}|$ contains a contribution from the
integral of $v_x$, and double periodicity shows up due to absolute value of the function. The
precession frequency as a function of the rotation frequency obtained from the auto-correlation
functions similar to that shown in Fig.22b, is presented in Fig.23. We were not able
to get such features from the PIV measurements due to restrictions in accuracy.\\

\section{Conclusions.}

We demonstrated that our system provides rather reliable information about both the phase and the
amplitude of the sound scattering signal by acquiring simultaneously 64 channels of the detector
array. The spatial and temporal information on the phase of the scattered signal allows us to get
values of circulation, vorticity, vortex location, and  vortex core radius. We verified quantitatively
the theoretical value for the slope of the proportionality coefficient, $K$, for a single rigid body
rotation vortex in a finite size cell. This method is rather comparable with the acoustic
time-reversal mirror (TRM) method \cite{fink1,fink2,fink3}. Instead of vorticity (or phase difference)
amplification due to number of crossing of the flow in the TRM method, in our method the
signal-to-noise ratio is amplified  by averaging complex wave functions over many pulses. So the
comparable resolution in the phase difference measurements
can be achieved.\\
At the same time in our approach  we can also use the amplitude of the scattering signal to
characterize the flow. The existing theories of the sound-flow interaction provide a relation between
the scattering signal and the structure functions of velocity and vorticity fields of a flow. However,
strong limitations imposed by the theories make an application of the sound scattering technique for
flow characterization rather restricted.  The theories consider the scattering signal produced by
 a sound emitter and obtained by a receiver of an infinite length in a far-field limit.\\
In the paper presented we studied both theoretically and experimentally  a possibility to obtain
reliable information about velocity and vorticity fields of a single either stationary or
time-dependent vortex flow by the acoustic scattering technique with a finite width sound beam of the
order of the vortex size and a finite size receiver taking a scattered signal rather close to the
scattering region. The experimental results and analysis are compared with the PIV measurements taken
simultaneously on the same vortex. From the theoretical side the main step in our analysis is the use
of the Huygens construction to obtain the acoustic scattering signal in a far field from the known
sound distribution in any intermediate plane. This reconstruction allows us to overcome rather
difficult experimental problem to obtain directly the scattering signal in a far-field. Such procedure
also helps us to realize that the only empirically relevant calculation of scattering is that of
 a confined flow, since an emitter and a detector define the flow perimeter.\\
 Another theoretical suggestion is the use of the beam function $\tilde{\Psi}_{rest}$, which describes the
 finite width beam and allows us to use the same formalism as for an infinite width beam.\\
The analysis of the experimental data based on the revised theory of sound scattering shows rather
good agreement between the sound scattering and PIV measurements, i.e. the sound scattering signal
obtained from the velocity field measured by PIV, coincides rather well with the scattering signal
obtained in the ultrasound experiment. The same can be said about the Fourier transform of the
vorticity obtained from the scattering via the construction suggested and that reproduced from the PIV
measurements. We also show that the peak value of the far field scattering signal is proportional to
the circulation or the angular speed of the core of a single vortex, and the angle distance between
the peaks is inversely proportional to the sound frequency. Further natural step is to use this
approach to a turbulent flow and to study the vorticity structure function
in a turbulent flow.\\

 This work is partially supported by Israel Science Foundation grant, by Binational US-Israel
 Foundation grant, and by the Minerva Center for Nonlinear Physics of Complex Systems.

\begin{figure}

\caption{Experimental set-up: (1) rotating plates, (2) light sheet and scattering plane, (3) PIV
camera with Schleimpflug optic arrangement, (4) acoustic emitter, (5) main signal pulse, (6) transmit
switch, (7) detector array, (8) 64 preamplifiers, (9) 64 lock-in amplifiers, (10) reference signal,
(11) sample and hold trigger, (12) acquisition cards (128 channels).}

\label{figa}
\end{figure}

\begin{figure}

\caption{Geometry of cell, flow and sound scattering. }

\label{figb}
\end{figure}

\begin{figure}

\caption{Numerical simulations of the Fourier transform of the velocity field of a single rigid body
rotation vortex flow $r_0=5$mm for a finite width beam, $d$,  compared to an infinite width beam: (a)
$d=120$mm, (b) $d=60$mm, (c) $d=20$mm. Solid line-a finite width beam, dash-dotted line-infinite width
beam.}

\label{figc}
\end{figure}

\begin{figure}

\caption{(a) Velocity field snapshot obtained by PIV for a single vortex flow created by co-rotating
rods of 25mm diameter; (b) Azimuthal velocity as a function of a cell radius extracted from PIV
measurements for a flow between rotating rods of 25mm diameter at $\Omega=71$rad/sec: black
dots-experimental points, solid line-the model with fitting parameters:  the angular speed
$\tilde\Omega=71$rad/sec and the core radius $r_0=6$mm. }

\label{figd}
\end{figure}

\begin{figure}

\caption{ Similar to the plot in Fig.4b but for a flow between rotating plates of 40mm diameter and
$\Omega=42$rad/sec. Solid line presents the fitted model with the core radius $r_0=14$mm }

\label{fige}
\end{figure}

\begin{figure}

\caption{ The same for a flow between rotating plates of 100mm diameter at $\Omega=23$rad/sec. Solid
line presents the fitted model with the core radius $r_0=42$mm. }

\label{figf}
\end{figure}

\begin{figure}

\caption{Spatial dependence of the phase shift by a flow between rotating plates of 40mm diameter at
$\Omega=42$rad/sec: stars-from sound measurements at 5.5MHz, circles-from PIV. }

\label{figg}
\end{figure}

\begin{figure}

\caption{Phase shift as a function of the scattering angle. The solid line is the fit, by which the
the phase shift slope is defined. The data are for a flow between the plates of 100mm diameter and
$\Omega=23$rad/sec with an emitter of 35mm length at $f=4$MHz.}

\label{figh}
\end{figure}

\begin{figure}

\caption{(a) Phase shift slope scaled by sound frequency, $(\partial \phi/\partial\theta)(1/f)$, as a
function of angular velocity for a flow between the plates of 100mm diameter and with emitter of 35mm
length; (b,c,d) Phase shift slope $(\partial \phi/\partial\theta)$ as a function of angular velocity
for a flow between the plates of 100mm diameter and with emitters the  of 143mm length, 10mm length,
and 3.5mm length, respectively, at various frequencies: squares-5.5MHz, diamonds-4.5MHz, dots-4.0MHz,
left-triangles-3.5MHz, up-triangles-2.5MHz, hexagons-2.0MHz, stars-1.5MHz, down-triangles-1.0MHz,
right-triangles-0.5MHz. }

\label{figi}
\end{figure}

\begin{figure}

\caption{The derivative of the phase shift slope on the rotation speed, $m$, as a function of sound
frequency for a flow between the plates of 100mm diameter for different size of emitters:
squares-3.5mm, dots-10mm, stars-35mm, triangles-143mm. Solid lines are the linear fits.}

\label{figj}
\end{figure}

\begin{figure}

\caption{Modulus of incident(a) and scattering(b) sound amplitude spatial distributions obtained by
the receiver (black dots) are presented. Solid lines are the calculations. The flow is produced
between two rotating plates of 100mm diameter, the emitter of 35mm length at 4MHz is used. }

\label{figk}
\end{figure}

\begin{figure}

\caption{The same as on the plots in Fig.11a,b but for the emitter of 125mm length at 5.5MHz.}

\label{figl}
\end{figure}

\begin{figure}

\caption{Evidence of side-lobs (diffraction pattern) obtained with emitter of 3.5mm length at 5.5MHz:
(a) incident, and (b) scattering sound amplitudes. }

\label{figm}
\end{figure}

\begin{figure}
\caption{Modulus of the structure function of the velocity and the beam function projected into a far
field for a flow between the plates of 40 mm diameter  and with the emitter of 125mm length. Dots are
the experimental data from sound scattering, solid line is calculation based on PIV measurements with
 the flow parameters $\tilde\Omega=42$rad/sec and $r_0=14$mm obtained from the fit, and 60 mm
 characteristic beam size is assumed due to the aperture.}

\label{fign}
\end{figure}

\begin{figure}

\caption{The same as on the plot in Fig.14 but for the emitter of 35mm length and the flow parameters
$\tilde\Omega=23$rad/sec and $r_0=42$mm obtained from the fit.}

\label{figo}
\end{figure}

\begin{figure}

\caption{Angular location of the structure function peak, $2\sin{(\theta_p/2)}$, as a function of the
reduced sound wave length, $(k_0d)^{-1}$. }

\label{figp}
\end{figure}

\begin{figure}

\caption{Value of the scaled structure function,
$|F\{v_x\tilde\Psi\}|g(r_0/d)d^{-0.8}(145/r_c)^{0.2}$, peak, $P$, as a function of the circulation,
$\Gamma$: stars- vortex flow with $r_0=6$mm created between two 25mm rods,
 60 mm beam width, and emitter T4; dots- vortex flow with $r_0=42$mm created between two 100mm plates,
 60 mm beam width, and emitter T4;
 squares-vortex flow with $r_0=14$mm created between two 40 mm plates, 60 mm beam width, and emitter T4;
 left-triangles-vortex flow with $r_0=42$mm created between two 100 mm plates, 35 mm beam width,
 and emitter T2.}

\label{figq}
\end{figure}

\begin{figure}

\caption{Reconstructed one-dimensional velocity field $(v_x)_{1D}$  according to the procedure
described in the text for 60 mm beam width, for a flow between two plates of 40 mm diameter and
$\Omega=42$ rad/sec, and an emitter T4 at frequency 5.5 MHz. Dashed line is an extrapolation based on
the flow model. }

\label{figr}
\end{figure}

\begin{figure}

\caption{(a) Modulus of the Fourier transform of the vorticity obtained from the data in Fig.18. (b)
Normalized spatial correlation function of the vorticity. Dashed line-sound scattering results, solid
line- calculations based on PIV velocity profile measurements as fig.5 (with extrapolation up to the
cell wall) with $r_0=14$mm and $\Omega=42$ rad/sec. }

\label{figt}
\end{figure}

\begin{figure}

\caption{Temporal dependence of the vortex location detected by PIV (dots in the upper set) and by
sound scattering (squares in the lower set) measurements.}

\label{figu}
\end{figure}

\begin{figure}

\caption{Auto-correlation functions for two sets of the data in Fig.20 as a function of the time
delay. Solid line- for sound scattering, dash-dotted line-for PIV.}

\label{figv}
\end{figure}

\begin{figure}

\caption{(a) Temporal dependence of the peak of the modulus of the velocity structure function for the
same measurements as in Fig.20. (b) Auto-correlation function for the data in (a) shows that the
vortex precession frequency is half of the peak frequency.}

\label{figw}
\end{figure}
\begin{figure}

\caption{Vortex precession frequency  obtained from the auto-correlation functions of the far-field
scattering peak time dependence (similar to the plot in Fig.22b), as a function of the frequency of
the plate rotation. The proportionality factor is $0.22\pm 0.02$.}

\label{figx}
\end{figure}
%\end{multicols}{2}

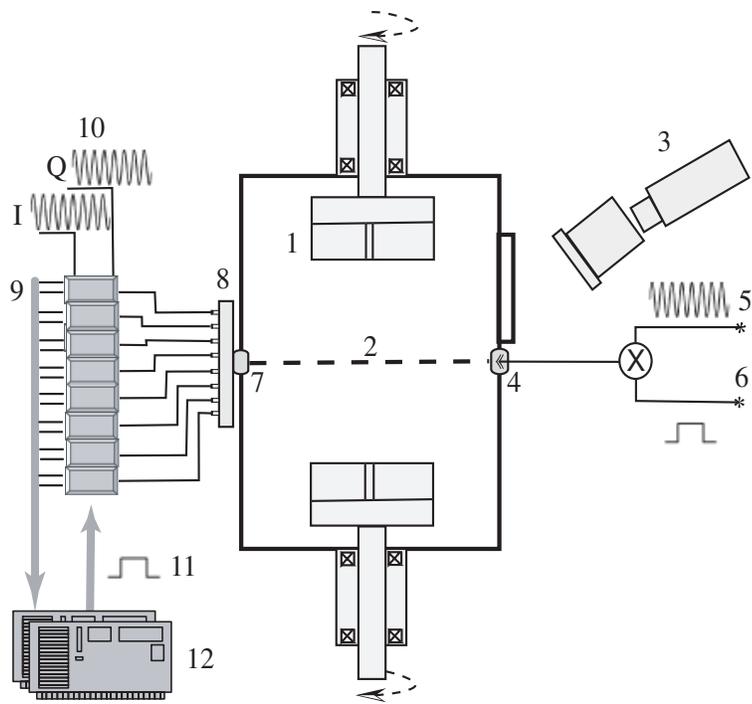
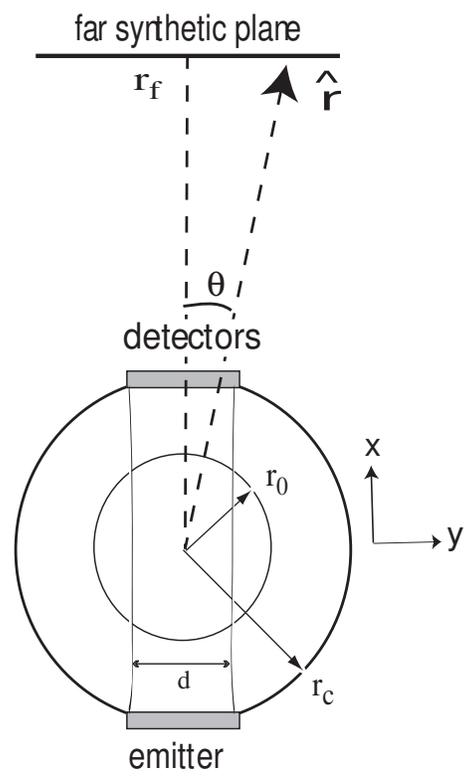
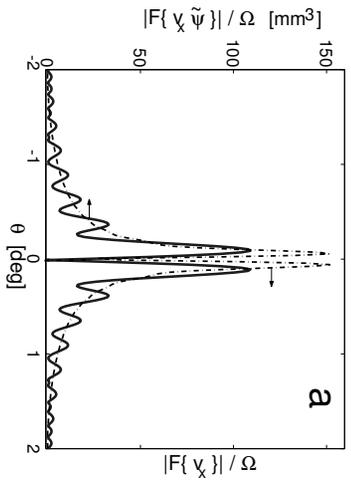
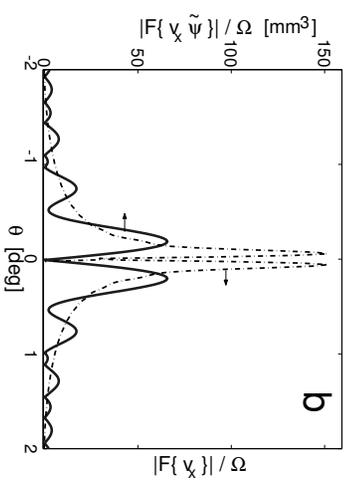
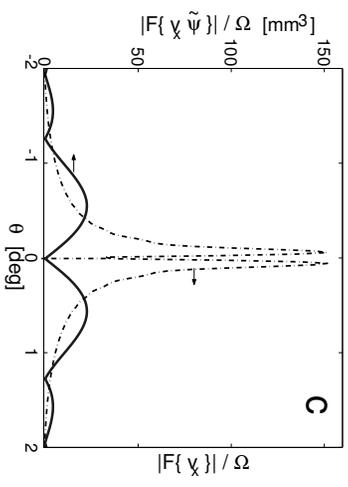
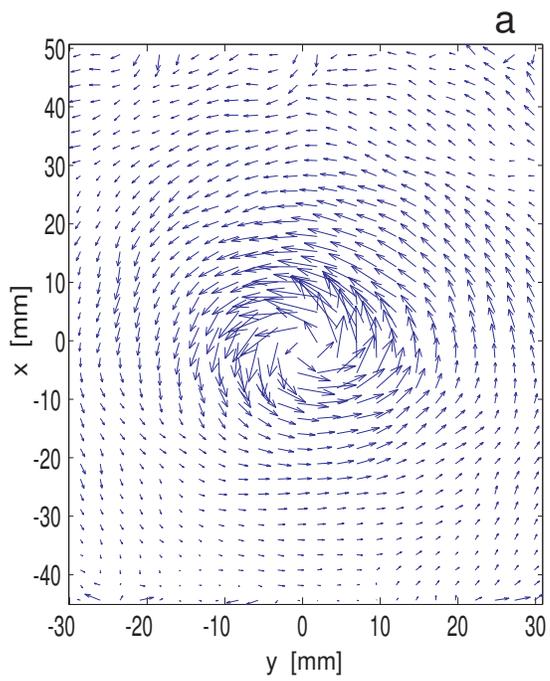
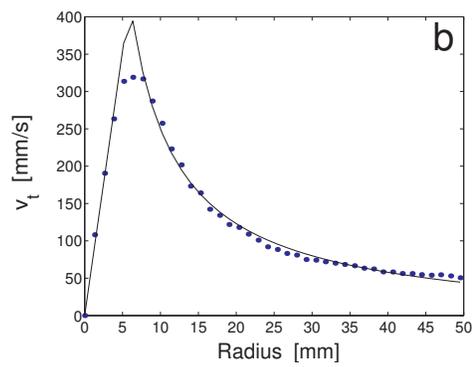
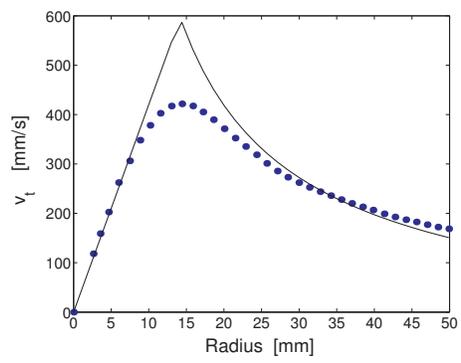
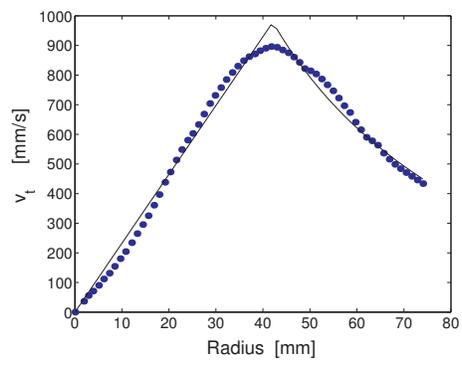
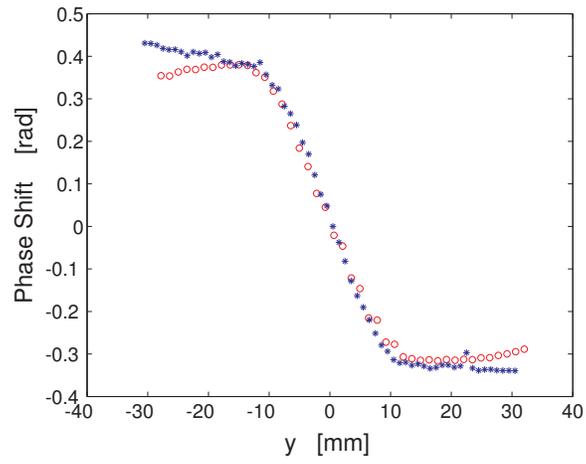
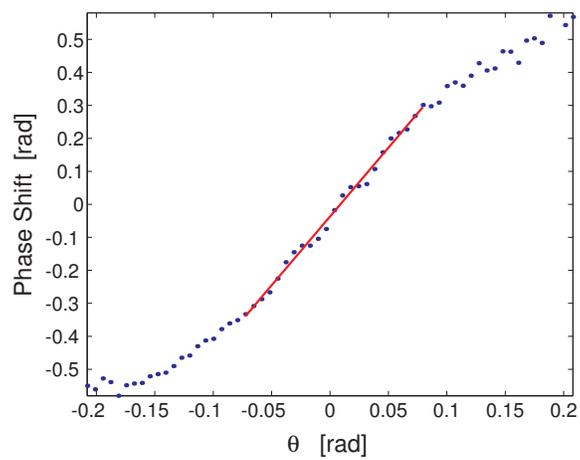
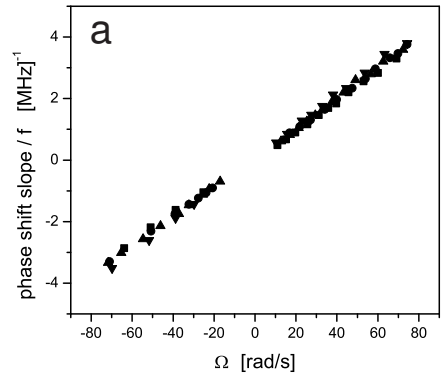
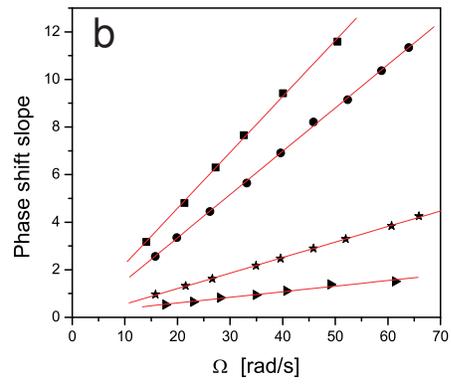
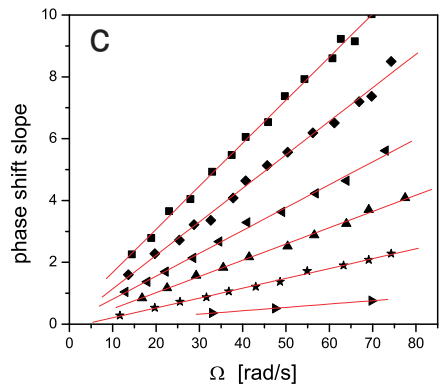
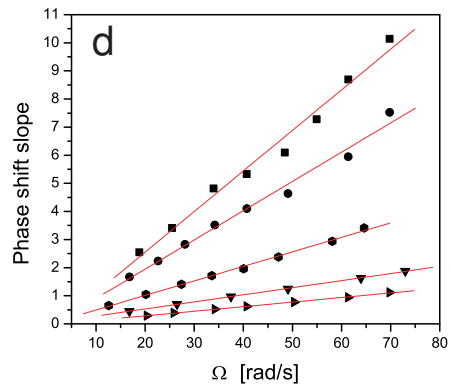
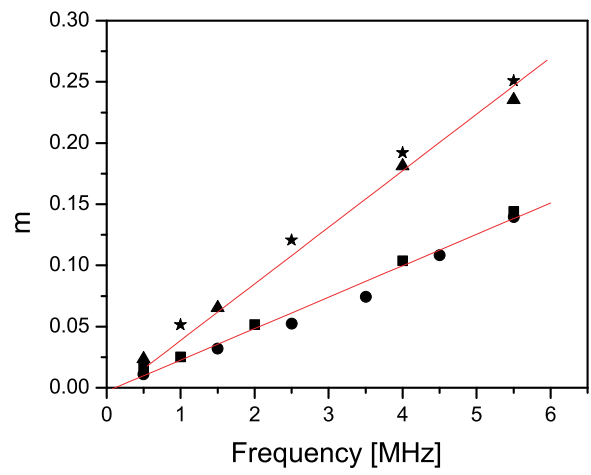
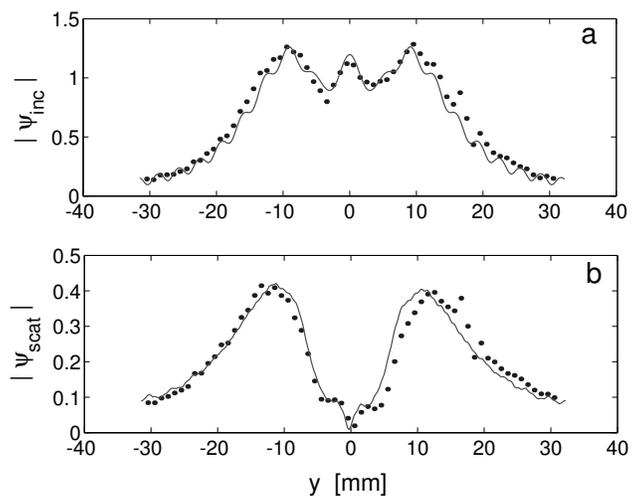
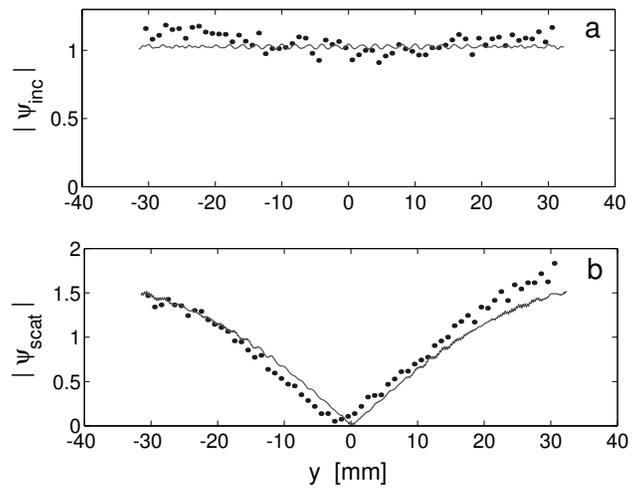
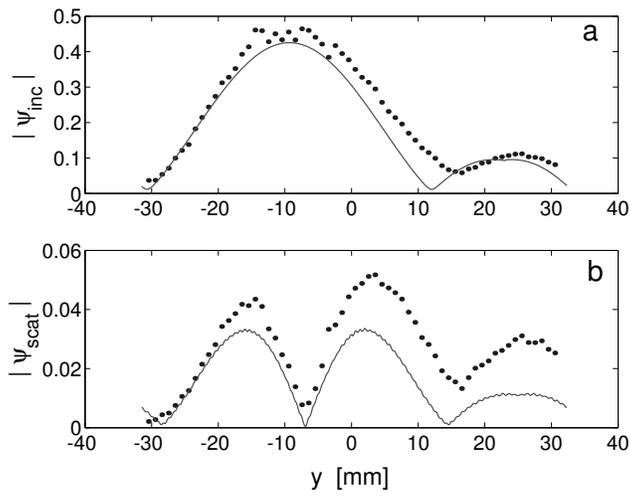
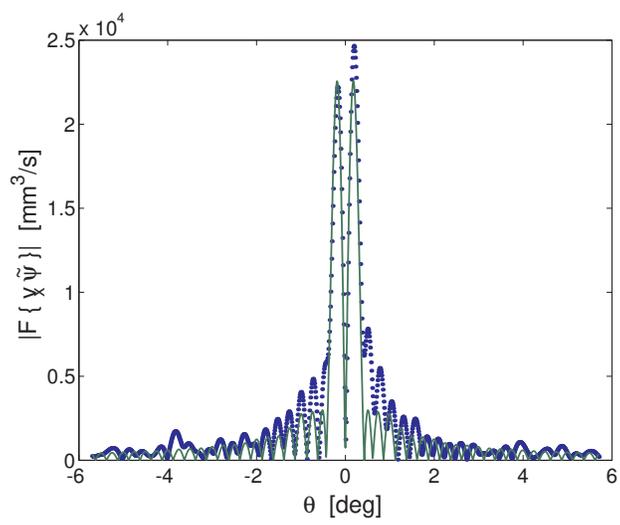
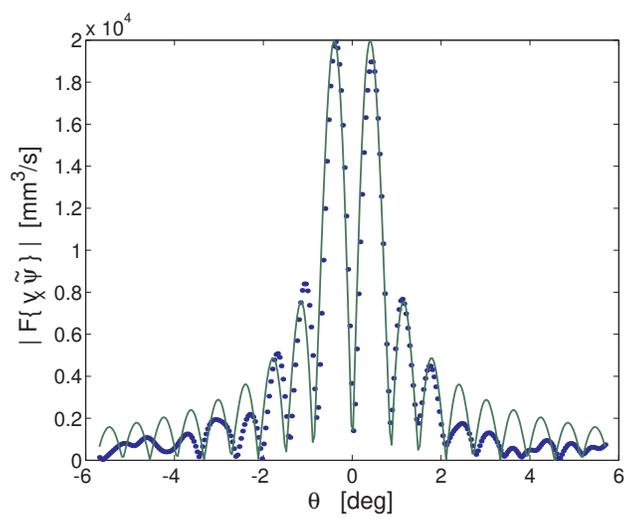
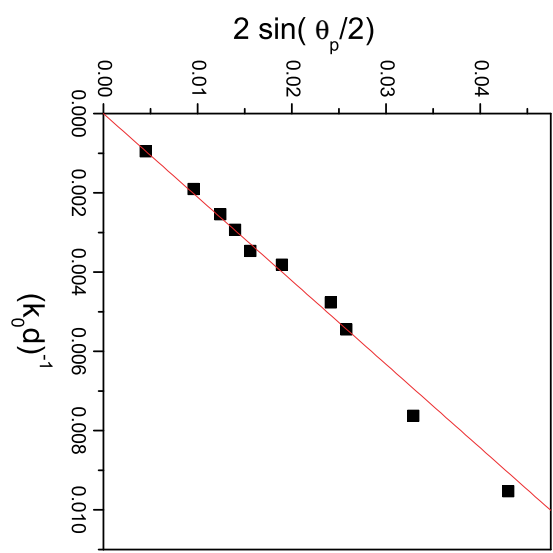
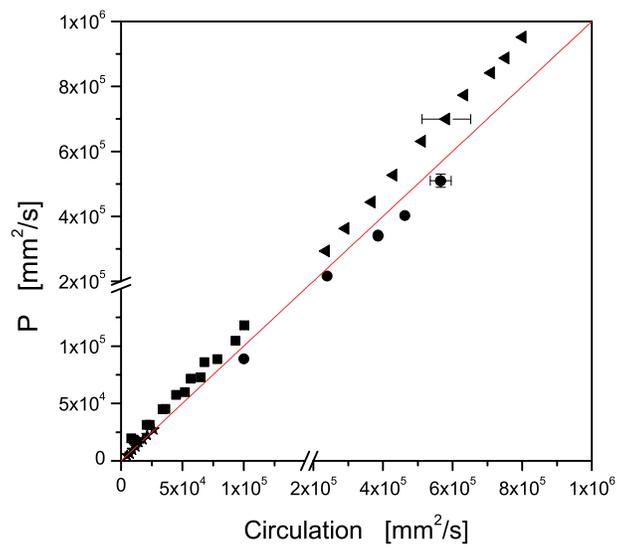
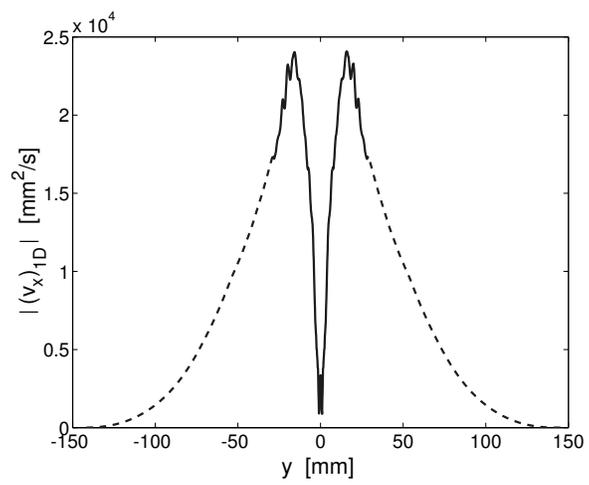
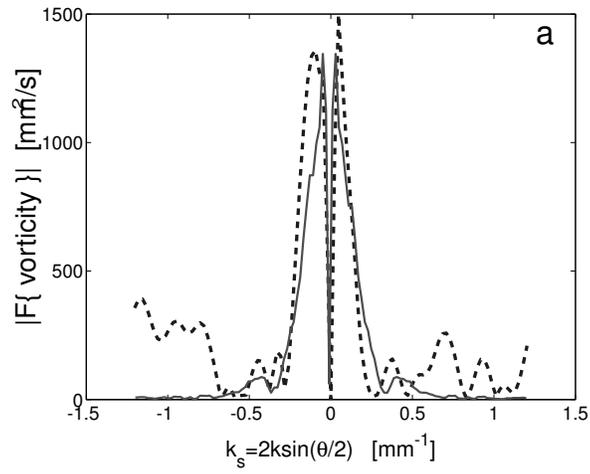
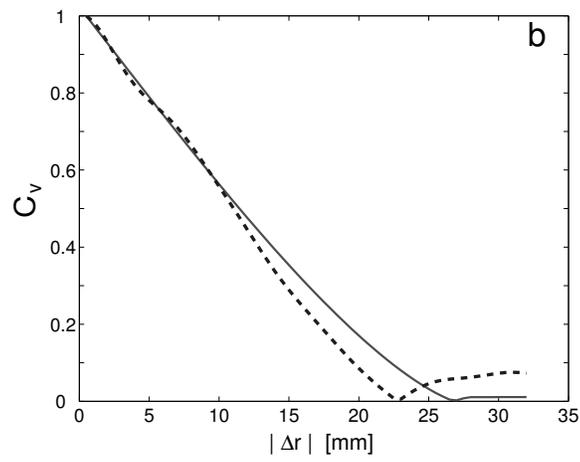
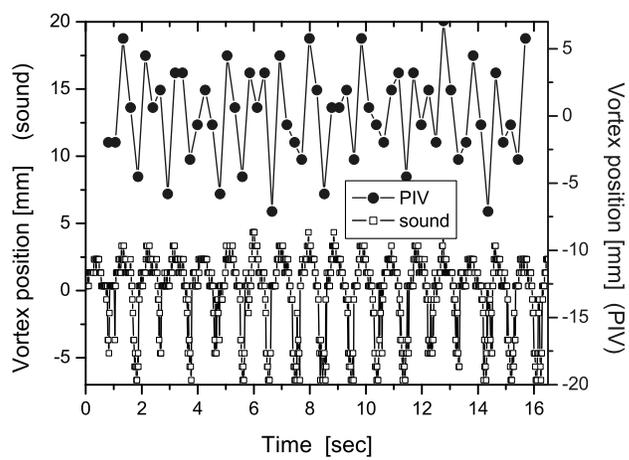
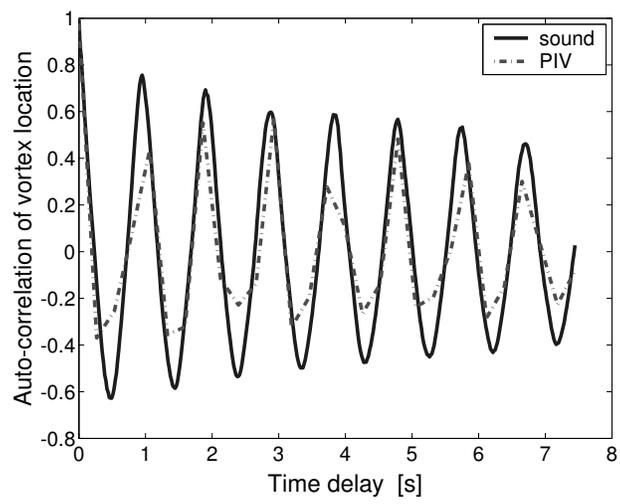
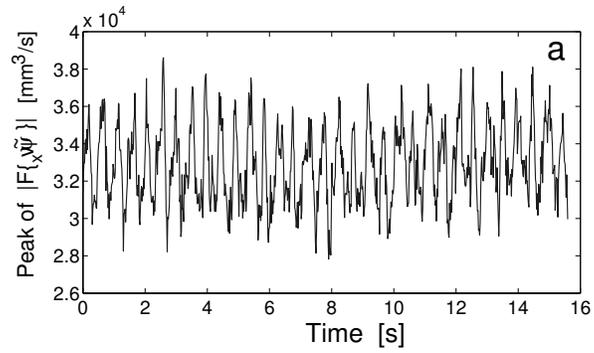
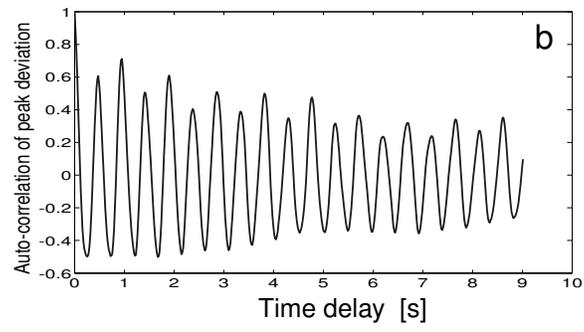
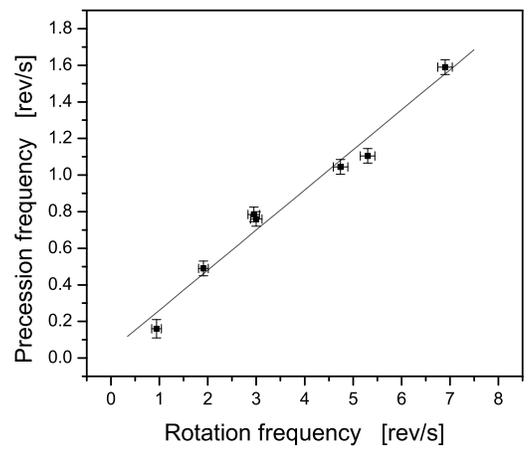
\end{document}